\documentclass[12pt]{article}

\usepackage{epsfig}

\usepackage{amssymb}
\usepackage{amsfonts}

\usepackage{color}
 
\def\be{\begin{equation}}
\def\ee{\end{equation}}
\def\ba{\begin{array}{c}}
\def\ea{\end{array}}

\newcommand{\bea}{\begin{eqnarray}}
\newcommand{\eea}{\end{eqnarray}}

\newcommand{\bbr}{\br\!\br}
\newcommand{\kkt}{\kt\!\kt}
\newcommand{\pbr}{\prec\!\!}
\newcommand{\pkt}{\!\!\succ\,\,}
\newcommand{\kt}{\rangle}
\newcommand{\br}{\langle}

\begin{document}

\begin{center}

{\Large \bf

Non-stationary non-Hermitian
``wrong-sign'' quantum oscillators
and their meaningful physical interpretation

}

\end{center}

\vspace{0.8cm}

\begin{center}

  {\bf Miloslav Znojil}$^{a,b}$

\end{center}

 $^{a}$  {Department of Physics, Faculty of
Science, University of Hradec Kr\'{a}lov\'{e}, Rokitansk\'{e}ho 62,
50003 Hradec Kr\'{a}lov\'{e},
 Czech Republic}

 $^{b}$
{The Czech Academy of Sciences,
 Nuclear Physics Institute,
 Hlavn\'{\i} 130,
250 68 \v{R}e\v{z}, Czech Republic,
{e-mail znojil@ujf.cas.cz}



\vspace{10mm}

\section*{Abstract}

In the framework of
quantum mechanics using quasi-Hermitian operators
the standard unitary evolution
of a non-stationary but still closed quantum system
is only properly described in the
non-Hermitian interaction picture (NIP).
In this formulation of the theory both the
states
and the observables
vary with time. A few aspects of
implementation of this picture are illustrated via
the
``wrong-sign'' quartic oscillators.
It is shown that
in contrast to the widespread belief,
both of the related
Schr\"{o}dinger-equation generators
$G(t)$
and the
Heisenberg-equation generators
$\Sigma(t)$
are
just auxiliary concepts.
Their
spectra are phenomenologically
irrelevant and, in general, complex.
It is argued that
only
the sum $H(t)=G(t)+\Sigma(t)$ of the latter operators
retains the standard physical meaning of the
instantaneous
energy of the unitary quantum
system in question.


\newpage

\section{Introduction}

In the $d-$dimensional and centrally symmetric
anharmonic-oscillator quantum Hamiltonian
 \be
 \mathfrak{h}^{(AHO)}(\lambda)=-\frac{1}{2}\,\triangle +\frac{1}{2}\,
 |x|^2+\lambda^2\,|x|^4\,,\ \ \ \ \ x \in \mathbb{R}^d
 \label{ahosc}
 \ee
the ``correct'' choice of the sign at the asymptotically dominant
and confining
${\cal O}(|x|^4)$ component of the interaction is known
to be responsible for the fact that the corresponding energy levels
$E_n(\lambda,\ell)$
form a discrete,  real and positive set
where $n,\ell=0,1,\ldots$, with $\ell$ being
the angular-momentum quantum number \cite{Fluegge}.

Incidentally, it is less widely known
that
the {\em same\,} discrete spectrum can be also interpreted as
corresponding to
a very different complex,
$\ell-$dependent  and, first of all,
``wrong-sign'' form
of the one-dimensional and asymptotically quartic
(and, obviously, manifestly non-Hermitian)
Hamiltonian-like operator
of Theorem Nr. 1
in paper~\cite{BG},
 \be
 \tilde{Q}(i\lambda,j)=-\frac{d^2}{dx^2}+\frac{j}{2}
 -{\rm i}j\lambda\,x+x^2-2{\rm i}\lambda\,x^3-\lambda^2\,x^4\,,
 \ \ \ \ \ j=2\ell+d-2\,.
 \label{tenles}
 \ee
The authors of the latter paper interpreted their isospectrality result
as a mere
mathematical curiosity \cite{VG}. Nevertheless, a few years later,
the perception of the wrong-sign models of the type (\ref{tenles}) has
changed: At present, it is widely accepted that there exist
many complex and
non-Hermitian Hamiltonians $H \neq H^\dagger$
possessing the real and discrete,
bound-state-mimicking
spectra \cite{Carl,ali,book}.

In retrospective, one can identify two
decisive ideas behind the
acceptability and
current acceptance
of the latter class of counterintuitive models
in the quantum model-building practice.
The emergence of the first idea dates back to
the brief letter \cite{BT} in which
Bender with Turbiner proposed that besides the
conventional ordinary differential
operators $H=-d^2/dx^2+V(x)$ in which the variable $x$
is real,
a consistent quantum theory might be also obtained
when
one performs an analytic continuation of the model
to a suitable complex curve of
$x \in {\cal C} \notin \mathbb{R}$.

In full strength such an idea has been
developed and presented, in 1998,
in the truly influential
letter by Bender and Boettcher \cite{BB}.
These authors proposed that
after analytic continuation
(during which $x$ becomes complex and
loses its observability status in general),
the ordinary differential operator $H=-d^2/dx^2+V(x)$
might still be accepted as
a consistent physical
Hamiltonian yielding the conventional, real and
observable
bound-state energy spectrum (see also review \cite{Carl}).

In the language of mathematics this means that
one has to
work, in general, with a ``complex coordinate'' $x$.
A few examples of such a choice are outlined
in section \ref{kap1} below.
In this context,
the second, equally important
support of the phenomenological acceptability
of the ordinary differential models
defined along a complex, ``unphysical'' curve of $x$
has been found in the ``quasi-Hermitian''
reformulation of the standard quantum mechanics
as provided by Scholtz et al \cite{Geyer}.
These authors, by their own words, ``established a general criterion
for a set of non-Hermitian operators to constitute a consistent
quantum mechanical system'', which would
``allow for the normal quantum-mechanical
interpretation'' and which would,
in the usual Hilbert space $L^2({\cal C})$,
``involve the construction of a metric'',
i.e., in our present notation, of
a suitable nontrivial and, in general,
Hamiltonian-dependent operator $\Theta=\Theta(H)\neq I$.

A more explicit account of the resulting amended
representation of the stationary quantum systems
(i.e., of the theory
which we will call non-Hermitian Schr\"{o}dinger picture, NSP)
may be found outlined in
section \ref{statek} below.
In its applications
the
analytic-continuation mathematics
appeared to yield many
new and unusual eligible
non-Hermitian but time-independent Hamiltonians $H\neq H(t)$
with real spectra (cf. the
review chapters in monograph \cite{Carlbook}).

Whenever people managed to construct
the metric (which appeared to be the case, in particular,
for the wrong-sign oscillators
of section \ref{kap1} below),
the new models appeared endowed also with
a meaningful physical (i.e., standard probabilistic) interpretation.
Naturally, the restriction of the above-mentioned NSP constructions
to the stationary wrong-sign potentials
is to be considered rather severe.
For this reason, Fring with Tenney \cite{FT}
opened the question of what happens when
these potentials become manifestly time-dependent.

In their search for inspiration, Fring with Tenney
restricted their attention to several
stationary wrong-sign
oscillators (this means, to the models listed
and discussed here in section \ref{kap1}).
Having proposed a
manifestly time-dependent generalization of these models,
they
arrived at a highly nontrivial and encouraging conclusion
that in spite of an enormous increase of the
complexity of the formalism (to be called here
non-Hermitian interaction picture, NIP),
they managed to reach
their ultimate goal of the closed-form construction
of all of the relevant operators.

In our present paper we intend to
complement the  Fring's and Tenney's
non-stationary and non-Hermitian wrong-sign-oscillator
results
by a few important methodical addenda.
As we already partially indicated,
our message will be preceded
by a compact review of the stationary wrong-sign models
of interest
(in section \ref{kap1})
and by a more detailed description and discussion of their
physical meaning in section \ref{statek}.
Then, the core of our
message will be presented in the methodically oriented
section \ref{tstatek}
(where we outline the basic features of the underlying NIP theory)
and in the application-oriented section \ref{ifutur}
where the Fring's and Tenney's construction
will be recalled and complemented by an outline of
an alternative approach.

A discussion and summary of our results will
finally be added in section \ref{sekcetreti}.


\section{Complex potentials with real energy levels\label{kap1}}

For a given, preselected and, what is important, stationary
and non-Hermitian Hamiltonian $H$
with the real spectrum and
with properties
$H \neq H(t)$ and
$H \neq H^\dagger$
in an unphysical, auxiliary Hilbert space $L^2({\cal C})$ {\it alias\,}
${\cal H}_{auxiliary}$,
the assignment of an alternative, amended physical
Hilbert space (endowed with a nontrivial {\it ad hoc\,} metric $\Theta$)
is of a decisive importance. It
makes such an operator quasi-Hermitian \cite{Dieudonne},
 \be
 H^\dagger\,\Theta=\Theta\,H\,,
 \ee
i.e., self-adjoint
with respect to the amended metric $\Theta\neq I$,
i.e., self-adjoint
in an amended, ``correct''
Hilbert space $L^2({\cal C},\Theta)$ {\it alias\,}
${\cal H}_{physical}$.

In such a context we felt motivated by the recent
developments of the
field in which the stationary-metric formalism
has been replaced by its
highly nontrivial non-stationary-metric generalization
(to be called non-Hermitian interaction picture, NIP,
in what follows).
In such an innovative framework,
even the ``anomalous'' potentials as sampled in Eq.~(\ref{tenles})
might be assigned
a meaningful quantum-theoretical interpretation
using the theory described, e.g.,
in the comprehensive reviews \cite{Carl,ali,book}.

\subsection{Constructions using perturbation theory}

In our present paper, we are going to pay detailed attention
to the possibility of
a consistent and meaningful quantum-theoretical interpretation
of the complex and manifestly non-Hermitian
Hamiltonians with the real bound-state spectra as
sampled by the latter ``wrong-sign'' example.
Our analysis will be restricted just to a few
unitary quantum models using the ``wrong-sign'' potentials
for which the
wave-function solutions of the corresponding
ordinary differential Schr\"{o}dinger equations
may be expected not to contradict a conventional bound-state
interpretation.

A consistent realization of the latter possibility
can rely upon the existing theory \cite{Carl,ali,book,Geyer}.
Temporarily we will assume
that all of our
Hamiltonians of interest remain stationary, $H \neq H(t)$.
This will enable us to explain, more easily, that such a
technical restriction
simplifies the meaningful physical interpretation of the systems and
leads to
a not too difficult  {\it ad hoc\,} reformulation
of quantum mechanics called non-Hermitian
or, better, quasi-Hermitian \cite{Dieudonne} quantum mechanics.

Initially, our interest in the problem was motivated
by the current
textbooks on quantum mechanics in which one can find
multiple illustrative examples of the
bound state energy spectra
of a particle moving in a real,
local and confining potential $V(\vec{x})$.
Besides the exactly solvable one-dimensional or
centrally symmetric harmonic-oscillator $V^{(HO)}(\vec{x})=|\vec{x}|^2$,
extensive attention is being paid also to its anharmonic
and, in particular, quartic
perturbations.
Although the meaningful
physical interpretation of the states in the
asymptotically confining (i.e.,
``correct-sign'') version
$V^{(AHO)}_{\lambda}(\vec{x})=|\vec{x}|^2+\lambda^2\,|\vec{x}|^4$
of the potential
remains standard,
unexpected problems
emerge
when one tries to construct the energy levels $E_n(\lambda,\ell)$ using
the Rayleigh-Schr\"{o}dinger perturbation series in the powers of $\lambda$.

During the history of such an approach
as summarized briefly in \cite{BG},
the problems caused by the divergence of perturbation series
were resolved (or, better,
efficiently circumvented) due to its Borel summability.
For our present purposes it is important that as a byproduct
of these results, a way appeared opened
towards a fully rigorous mathematical understanding of the
spectra of the wrong-sign operators as sampled by Eq.~(\ref{tenles})
above.

Initially, and not too surprisingly, people did not
call the corresponding Hamiltonian-like operators Hamiltonians.
One of the reasons was that
the normalizability of its eigenstates
appeared to be merely guaranteed by a
fine-tuned interference between the separate
components of the
asymptotically repulsive ``potential''.
Thus, people were only speaking, in such a case, about
``unstable oscillators'' \cite{BG}.

The paradigm has only been changed
when Bender with Boettcher  \cite{BB}
turned attention to the undeniable relevance of the
parity times time reversal {\it alias\,}
${\cal PT}-$operator symmetry of the models with real spectra.
On these grounds they
decided to advocate the anomalous non-Hermitian
models with real spectra as,
in some sense, phenomenologically acceptable.

\subsection{Constructions using ${\cal PT}-$symmetry}

In the traditional textbooks on quantum mechanics
the observables are assumed self-adjoint
(see, e.g., \cite{Messiah}).
Precisely this constraint has been
criticized by Bender with Boettcher as too formal \cite{BB}.
These authors initiated the development of
an innovative but still
conceptually tenable quantum
theory (a.k.a. ${\cal PT}-$symmetric quantum mechanics, PTQM,
see, e.g., its detailed review \cite{Carl}).
In this approach
(which may also be called non-Hermitian Schr\"{o}dinger picture, NSP)
the
bound states
were conjectured represented by the
eigenvectors $|\psi(t)\kt \in {\cal H}^{(BB)}$ of an unusual,
stationary (or quasi-stationary \cite{ali}) but
manifestly non-Hermitian
Hamiltonian $H \neq H^\dagger$.

The impact of the Bender's and Boettcher's
conjecture has been enhanced by their
explicit demonstration that the
hypothetical new family of the eligible non-Hermitian
candidates for the quantum Hamiltonians
of unitary systems
may even contain a fairly rich subfamily of
Hamiltonians with the conventional
kinetic-energy plus potential-energy structure. In
their explicit illustrative example
$H^{(BB)} = -d^2/dz^2+V(z)$ the potential
has been chosen complex but still local and
analytic,
 \be
 V(z)=V^{(BB)}(z,\delta)= \lambda^2\,z^2\,({\rm i}z)^\delta\,,
 \ \ \ \ \delta \geq  0\,.
 \label{model}
 \ee
Strictly speaking,
up to the exceptional harmonic-oscillator special case with $\delta=0$,
the intuitive
potential-energy treatment
of such an interaction
would not be appropriate. For two reasons. The first,
less important one is that
even the coupling constant is complex in general. Hence,
unless $\delta=0,2,4,\ldots$,
the interaction itself would
not be too easily realizable in the laboratory
even when one succeeds in keeping
the coordinate itself
observable, $z \in \mathbb{R}$.

The second counterintuitive feature of the
${\cal PT}-$symmetric interaction
(\ref{model}) is that its definition admitting
the real and observable
coordinate $z \in \mathbb{R}\,$ {\em and\,}
the real and observable
bound-state energies $E_0, E_1, \ldots$
only appeared phenomenologically consistent
in a finite interval $0\leq \delta < 2$ of the exponents.
Otherwise, at negative $\delta >-1$
a high-lying part of the energy spectrum
has been all found complex
(and, hence, unobservable)
\cite{BB}.
After all, the whole spectrum
becomes empty in the Herbst's limit $\delta \to -1$ \cite{Herbst}.
In parallel, at any larger $\delta \geq 2$ the correct asymptotic values of
the variable $z$
had to be chosen complex
and, hence, unobservable (cf., e.g., \cite{BB} or \cite{BBwell}
for explanation).

\subsection{The Jones' and Mateo's wrong-sign model with $\delta=2$}

In the context of preceding paragraph a special role is played by the
Bender's and Boettcher's Hamiltonians
with $\delta = 2$.
A fairly
persuasive illustration of the phenomenological as well as formal appeal of
such a remarkable special case
has been provided by Jones and Mateo \cite{JM}.
In their analysis they
considered the Hamiltonian
 \be
 H^{(JM)}(\lambda)=-\frac{d^2}{dz^2} - \lambda^2\,z^4\,
 \label{haha}
 \ee
which is by far the simplest quartic wrong-sign oscillator
with real spectrum.

The Jones' and Mateo's proof of the reality of the spectrum
of their model was constructive.
First, they specified the explicit form
of the complex contour of the ``coordinate'' $z$,
 \be
 z = z(x) = -2{\rm i} \sqrt{1+{\rm i}x}\,,\ \ \ \ \ x \in \mathbb{R}\,.
 \label{kont}
 \ee
In terms of the new real coordinate $x$ the original Hamiltonian (\ref{haha})
acquires the form of an ordinary differential operator $H=H_0+H_1$ (cf. Eq. Nr. (11) in
{\it loc. cit.}) composed of a Hermitian part $H_0=p^2-p/2+16\lambda^2(x^2-1)$
and an anti-Hermitian component $H_1={\rm i}(xp^2+p^2\,x)/2 -32{\rm i} \lambda^2\,x
$ (cf. Eq. Nr. (13) in
{\it loc. cit.}).


\begin{figure}[h]                     
\begin{center}                         
\epsfig{file=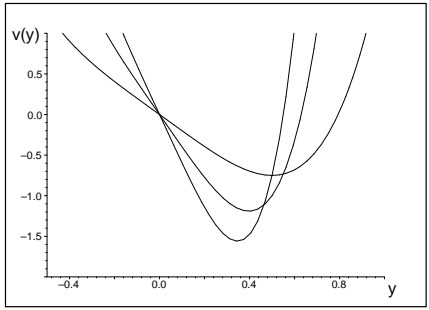,angle=270,width=0.32\textwidth}
\end{center}                         
\vspace{-2mm} \caption{The shape of potential  (\ref{pojdte})
at $\lambda=1,2$ and $3$.
 \label{ureone}}
\end{figure}

Due to the simplicity of the
Jones' and Mateo's ``wrong sign''
model it appeared feasible
to prove that the ``anomalous'' operator $H^{(JM)}(\lambda)$
is in fact fairly well behaved and, first of all, strictly isospectral
to another, conventional and self-adjoint
Hamiltonian
 \be
 \widetilde{\mathfrak{h}}^{(JM)}(\lambda)=
 -\frac{d^2}{dy^2} - 2\,\lambda\,y
 +4\,\lambda^2\,y^4\,.
 \label{ouha}
 \ee
Here, the potential is safely confining and smooth (cf. its samples in
Figure \ref{ureone}).
Near its minimum $\mathfrak{v}(y_{\min})=-3\lambda^{2/3}/4\,$
at $\,y_{\min} = 1/(2\,\lambda^{1/3})\,$
this potential is reasonably well approximated
by the exactly solvable harmonic-oscillator well,
 \be
 \mathfrak{v}(y)=
   - 2\,\lambda\,y
 +4\,\lambda^2\,y^4 \approx
 \mathfrak{v}(y_{\min}) + \omega^2(y-y_{\min})^2 + {\cal O}[(y-y_{\min})^3]\,,
  \label{pojdte}
 \ee
with
$\omega^2=\mathfrak{v}''(y_{\min})/2=24\,\lambda^2\,y_{\min}^2$.
With the growth of $\lambda$
the minimum decreases and
the well
becomes narrower. Despite the decrease of the minimum
the value $E_0 \approx
\mathfrak{v}(y_{\min})+\sqrt{6}\,\lambda^{2/3}$
of the ground-state energy itself
remains positive and growing because $\sqrt{6}-3/4>0$.


\subsection{The Buslaev's and Grecchi's anharmonic wrong-sign model}


In the above-mentioned perturbation-theory-based analyses of the
conventional ``correct-sign'' anharmonic-oscillator
bound-state energies $E_{n}(\lambda,\ell)$
the mathematics becomes most interesting
when one imagines that these energies
can be represented by the Borel sum of the conventional
(and, in this case, divergent)
Rayleigh-Schr\"{o}dinger perturbation series \cite{Fluegge}.
This opens the way towards the perception
of the spectrum in the language of
analytic functions.

In this spirit,
Buslaev with Grecchi \cite{BG}
recalled the the radial component of the
conventional anharmonic-oscillator
Hamiltonian (\ref{ahosc})
and
emphasized that after
a rotation of the coupling in complex plane ($\lambda \to {\rm i}\lambda$)
one obtains a wrong-sign anharmonicity in infinity and
a new type of singularity in the origin..
After its
regularization
(via a complex shift of
$r \to r_\epsilon(x)=x-{\rm i}\epsilon$)
they obtained a new Hamiltonian of the form
 \be
 H^{(BG)}=H_\epsilon({\rm i}\lambda,j)=
 \frac{1}{2}\,
 \left (-\frac{d^2}{dx^2}+
 \frac{j^2-1}{4\,r_\epsilon^2(x)}+r_\epsilon^2(x)
 \right )-\lambda^2\,r_\epsilon^4(x)\,.
 \label{BUGR}
 \ee
This operator
is manifestly non-Hermitian but,
in the Buslaev's and Grecchi's terminology,
it is still $TP-$symmetric in $L^2(\mathbb{R})$,
with $P$ representing
parity, and with $T$ denoting the operator of complex conjugation
(cf. remark Nr. 4 in \cite{BG}).

On this background these authors were able to prove that
their oscillator (\ref{BUGR})
is isospectral
with the fully conventional
and safely self-adjoint and confining double-well
Hamiltonian
 \be
 \mathfrak{h}^{(BG)}=Q(\lambda,j)=-\frac{d^2}{dy^2}+j/2-j\,\lambda\,y+
 y^2
 -2\,\lambda\,y^3
 +\lambda^2\,y^4\,
 \label{UGR}
 \ee
(cf. Theorem Nr. 5 in {\it loc. cit.}).
This implies that
the shared spectrum of
{\em both\,} of these Hamiltonians is real,
discrete and bounded from below.

One might conclude that
in contrast to the widespread
use of the strongly misleading nickname ``unstable oscillator'' \cite{FT},
the wrong-sign anharmonic oscillator (\ref{BUGR}) may be tentatively
interpreted as representing a closed quantum system
which is unitary and stable.
Naturally, it would only be necessary to reformulate
the conventional quantum mechanics of textbooks.
Precisely such
a reformulation of the abstract theory has been provided
by Scholtz et al \cite{Geyer}.
Let us now recall and outline its basics.

\section{Hiddenly Hermitian theory in stationary regime\label{statek}}


In many conventional introductions to quantum mechanics of unitary systems
the states are represented by the time-dependent ket-vector
elements $|\psi(t)\pkt$ of a suitable physical Hilbert space ${\cal L}$
while the operators $\mathfrak{q}$ representing  observables are usually
kept time-independent,
$\mathfrak{q} \neq \mathfrak{q}(t)$.
Under these assumptions
characterizing
quantum theory formulated in
the so called ``Schr\"{o}dinger picture'' (SP, \cite{Messiah})
the predictions of the results of measurements
are given in terms of matrix elements
$\pbr \psi(t)|\mathfrak{q}|\psi(t)\pkt$.
Thus, in applications one only has to solve
Schr\"{o}dinger equation
 \be
 {\rm i}\frac{d}{dt}\,|\psi(t)\pkt=
  {}{\mathfrak{h}}\,|\psi(t)\pkt
 \label{sekv}
 \ee
while
using the Hamiltonian which
must be self-adjoint in ${\cal L}$, ${\mathfrak{h}}=
{\mathfrak{h}}^\dagger$.

\subsection{Non-Hermitian Schr\"{o}dinger representation}

The situation becomes different when a candidate $H$ for the Hamiltonian
is non-Hermitian. Its standard quantum-mechanical interpretation can be
then established in a way described, in 1992, by Scholtz et al \cite{Geyer}.
The basic idea of the reformulation of the
quantum mechanics of unitary systems
in non-Hermitian Schr\"{o}dinger picture (NSP, see \cite{Geyer} and also the
more recent reviews \cite{ali,book})
can be seen in a replacement of the single physical Hilbert space of states
(say, ${\cal L}$)
by the two mutually non-equivalent Hilbert spaces
(say, ${\cal H}_{physical}$ and  ${\cal H}_{auxiliary}$).
In such a setting one has to generalize the
conventional Schr\"{o}dinger picture (SP) and to replace the underlying usual
Schr\"{o}dinger
equation (\ref{sekv})
by an appropriate modification.

The basic idea behind the generalization is that a state of the system
(marked, say, by the Greek letter $\psi$ and varying with time,
$\psi=\psi(t)$) is perceived not only as a ket-vector
element of the
conventional textbook space ${\cal L}$ (to be written here as
a ``curly'' ket $|\psi(t)\pkt$)
but also as a (shared)
element $|\psi(t)\kt$ of both of the alternative Hilbert spaces
${\cal H}_{physical/auxiliary}$  (with the state
marked, in both spaces, by the
same and
most common version of the ket-symbol).

First of all, the formal SP - NSP equivalence of the description
may be given the form of the one-to-one correspondence
between the representations,
 \be
 |\psi^{(textbook)}\pkt=\Omega_{}\,|\psi^{(auxiliary)}\kt
 \,.
  \label{lumapo}
 \ee
Using the concept of the (by definition, nontrivial \cite{Geyer})
metric operator $\Theta=
 \Omega^\dagger\,\Omega\neq I$, we may immediately
introduce the bra vectors denoted as $\pbr \psi(t)|$ in ${\cal L}$,
as $\br \psi(t)|$ in ${\cal H}_{auxiliary}$, and
as $\bbr \psi(t)|$ in ${\cal H}_{physical}$.
Now, once we recall the mapping
(\ref{lumapo}) we may
simply avoid the related ambiguities via a
reinterpretation of ${\cal H}_{physical}$
as a mere modified version of space ${\cal H}_{auxiliary}$
in which the inner product is amended,
 \be
 \br \psi
 |\chi \kt \to \br \psi|\Theta |\chi \kt \
  \ee
or, equivalently, in which one
simply represents $\bbr \psi(t)|=\br \psi(t)|\Theta$
and $|\psi(t)\kkt = \Theta|\psi(t)\kt$.

On can notice that such a notation convention is compact,
efficient and consistent.
For example, having an observable
represented by an operator $\mathfrak{q}$ in ${\cal L}$
(where it has to be self-adjoint),
we can just work with its isospectral avatar in ${\cal H}_{auxiliary}$,
 \be
 Q_{}= \Omega_{}^{(-1)}\,
 \mathfrak{q}_{}\,\Omega_{} \neq Q_{}^\dagger\,.
 \label{dadag}
 \ee
Nevertheless, it is not necessary to
emphasize the self-adjointness of $Q_{}$ in ${\cal H}_{physical}$
because it is sufficient to stay working in ${\cal H}_{auxiliary}$
where such a requirement acquires the metric-dependent form
 \be
 Q^\dagger\,\Theta=\Theta\,Q\,.
 \label{[15]}
 \ee
This property of the operator $Q$ imposed in the
preferred but unphysical working Hilbert space
${\cal H}_{auxiliary}$ is called,
in mathematics, quasi-Hermiticity \cite{Geyer,Dieudonne}
{\it alias\,} $\Theta-$pseudo-Hermiticity \cite{ali}.

Using our present notation conventions we may summarize that
the predictions mediated by the SP and NSP calculations
are
indistinguishable,
in essence, due to the coincidence of matrix elements
 \be
 \pbr \psi^{}(t)\,|\mathfrak{q}_{}\, |\psi^{}(t)
 \pkt=
 \bbr \psi^{}(t)\,|Q_{}\,
 |\psi^{}(t)
 \kt\,.
 \label{dysMEAS}
  \ee
Thus, once we restrict our attention
just to the NSP framework,
it is sufficient to
find a (time-independent) solution $\Theta=\Theta(Q)$
of Eq.~(\ref{[15]}). Then, for an evaluation of
predictions (\ref{dysMEAS}) one just needs
to know the time-dependent solution
$|\psi(t)\kt$ of
the
NSP evolution equation,
 \be
 {\rm i} \frac{\partial}{\partial t} \,|\psi^{}(t)\kt
 =H_{}\,|\psi^{}(t)\kt\,.
 \label{SETdys}
 \ee
Obviously, all of the calculations may
be performed, exclusively, in the preferred
``mathematical'' Hilbert space ${\cal H}_{auxiliary}$.

\subsection{The case of stationary wrong-sign potentials}

The main problem addressed in the Jones' and Mateo's paper \cite{JM}
was the construction of the inner-product metric $\Theta^{(JM)}$
which would make their wrong-sign quartic Hamiltonian (\ref{haha})
quasi-Hermitian in ${\cal H}_{auxiliary}$. The essence of their
discovery was that in the corresponding constraint (\ref{[15]})
with $Q=H^{(JM)}(\lambda)$, viz., in relation
 \be
 \left [H^{(JM)}\right ]^\dagger(\lambda)\,\Theta^{(JM)}=
 \Theta^{(JM)}\,H^{(JM)}(\lambda)
 \ee
the insertion of a suitable formal
power-series ansatz
for $\Theta^{(JM)}$
led to a termination of the infinite series and to a closed
and compact inner-product metric
 \be
 \Theta^{(JM)}(\lambda)= \exp \left [
 p^3/(48\,\lambda^2)-2\,p
 \right ]
 \label{[19]}
 \ee
(cf. formula Nr. 16 in {\it loc. cit.}).
In this manner,  Jones with Mateo  obtained the desirable
meaningful physical interpretation of their wrong-sign quartic model
in which all of the observable quantities have to be represented by
the operators $Q^{(JM)}$ which have to obey the quasi-Hermiticity
constraint (\ref{[15]}) in ${\cal H}_{auxiliary}=L^2(\mathbb{R})$.

A formally analogous constructive NSP recipe
can be also applied to the other
wrong-sign anharmonic oscillators and/or to their
${\cal PT}-$symmetric stationary alternatives.
In the Buslaev's and Grecchi's model \cite{BG}, in particular,
the presence of a more complicated components in the interaction
made also the $\Omega-$mediated reconstruction
of the isospectral self-adjoint Hamiltonian $\mathfrak{h}^{(BG)}$
of Eq.~(\ref{UGR}) conceptually fully analogous
but, in
the purely technical terms,
perceivably more complicated,
with a central role still played by an interchange $p \leftrightarrow x$
mediated by Fourier transformation.

\section{Quantum theory in non-stationary dynamical regime\label{tstatek}}

Briefly, the transition from the conventional, Hermitian version
of quantum mechanics (represented
in Schr\"{o}dinger picture \cite{Messiah})
to its non-Hermitian (or, better, quasi-Hermitian)
generalization can be characterized as a
separation and transfer of the
two different roles played by the single textbook Hilbert space
${\cal L}={\cal H}^{(T)}$ to the two non-equivalent Hilbert spaces,
viz.,
to the ``friendly'' space ${\cal H}^{(F)}$ and
to the ``physical'' space ${\cal H}^{(P)}$.

\subsection{Evolution equations for
states\label{amo}}


After one
decides to relax the auxiliary technical stationarity assumption
and after one admits the manifestly time-dependent
Dyson maps $\Omega=\Omega(t)$ and metrics
$\Theta=\Omega^\dagger(t)\,\Omega(t)=\Theta(t)$,
an enhancement of the flexibility of the formalism
becomes accompanied by an increase of the complexity of the
theory.

Naturally, the basic idea, i.e., the
use of the (this time, time-dependent) Dyson mapping
 \be
 |\psi^{}(t)\pkt=\Omega(t)\,|\psi^{}(t)\kt
 \in {\cal H}^{(T)}\,,
 \ \ \ \ \ \ |\psi^{}(t)\kt \in {\cal H}^{(F)}\,
 \label{mapoge}
 \ee
remains unchanged. Nevertheless, its consequences become
less straightforward. First of all, an
easy observation is that the self-adjoint Hamiltonian
$\mathfrak{h}_{(SP)}$ or $\mathfrak{h}_{(SP)}(t)$ of
textbooks still becomes replaced by its
isospectral ``quasi-Hermitian-Hamiltonian'' image
 \be
 H_{}(t)= \Omega_{}^{(-1)}(t)\,
 \mathfrak{h}_{(SP)}(t)\,\Omega_{}(t)\,
 \label{udagobs}
 \ee
acting in ${\cal H}^{(F/P)}$ or, more precisely, acting and
self-adjoint in ${\cal H}^{(P)}$ and acting and
non-Hermitian in ${\cal H}^{(F)}$ while, in a more appropriate
language of review \cite{Geyer},
compatible with the constraint
 \be
  H^\dagger(t)\, \Theta(t)=
  \Theta^{}(t)\,H_{}(t)
 \,,
 \ \ \ \
 \Theta(t)=\Omega^\dagger(t)\,\Omega(t)\,
 \label{dagobs}
 \ee
{\it alias\,}
quasi-Hermitian in ${\cal H}^{(F)}$.

On these grounds is is not too difficult to
recall the NIP formalism of papers \cite{timedep,SIGMA} and
re-write the
SP Schr\"{o}dinger Eq.~(\ref{sekv}) of textbooks in its upgraded
NIP form in ${\cal H}^{(F)}$,
 \be
 {\rm i} \frac{\partial}{\partial t}
 \,|\psi^{}(t)\kt=G_{}(t)\,|\psi^{}(t)\kt\,,
 \ \ \  \ G(t)=H(t)-\Sigma(t)
  \label{SEFip}
 \ee
where the symbol
 \be
 \Sigma(t)={\rm i} \Omega^{-1}(t)\,\dot{\Omega}(t)\,,
 \ \ \ \ \
 \dot{\Omega}(t)=
 \frac{d}{dt} \,
 \Omega(t)\,
 \label{defsig}
 \ee
denotes the so called Coriolis force -- see also \cite{NIP}
for details.

In the latter review \cite{NIP} we noticed
that in the non-stationary NIP setting the time-dependence of the metric
makes its explicit construction too costly.
For this reason we strongly recommended
to avoid its use and, whenever possible,
to replace it by the
use of the ``ketkets'',
i.e., of the alternative ket vectors
$ |\psi(t)\kkt = \Theta(t)\,|\psi(t)\kt$
in ${\cal H}^{(F)}$.

From the purely pragmatic point of view this
can lead to important simplifications
because one can easily verify that
 \be
 |\psi^{}(t)\pkt
  =\left [\Omega^\dagger(t)\right ]^{-1}|\psi(t)\kkt
 \in {\cal H}^{(T)}\,.
 \label{dumapo}
 \ee
The definition of the ketkets implies that
they may be defined as solutions of a ``second''
Schr\"{o}dinger-like equation in ${\cal H}^{(F)}$,
 \be
 {\rm i} \frac{\partial}{\partial t} \,|\psi(t)\kkt
 =G^\dagger(t)\,|\psi(t)\kkt\,.
  \label{d2.3}
 \ee
One can now add the two observations which are important.
The first one is that in the NIP framework we can still
consider any preselected
(and, naturally, quasi-Hermitian) operator (say, $Q(t)$)
representing an observable,
 \be
 Q_{}(t)= \Omega_{}^{(-1)}(t)\,
 \mathfrak{q}_{(SP)}(t)\,\Omega_{}(t)\,
 \label{redadag}
 \ee
i.e., quasi-Hermitian,
 \be
 Q_{}^\dagger(t)\,\Theta(t)=\Theta(t)\,Q(t)\,.
 \label{nenene}
 \ee
The second observation is that
once we
are given $Q(t)$
and use
and solve the second Schr\"{o}dinger-like equation,
we are
immediately able to
evaluate the matrix elements of interest,
\be
 \bbr \psi^{}(t_f)\,|Q_{}(t_f)\, |\psi^{}(t_f) \kt\,
 \label{redysMEAS}
  \ee
i.e., in other words, we can in fact
predict the results of the measurements
without the explicit construction of the metric.

\subsection{Physics behind the equations}

Using our present notation we read,
in Theorem Nr. 2 of review \cite{ali},
that ``if the time evolution \ldots is unitary
and $G(t)$ is an observable, \ldots then the metric
\ldots does not depend on time, i.e., there must exist a
time-independent operator $\Theta$''.
The statement has been based on a tacit assumption that
``the time-evolution of the system \ldots is determined
by the Schr\"{o}dinger equation'' (i.e.,
in our present paper, by Eq.~(\ref{SEFip})).

In the preceding subsection we explained
that such an assumption is only acceptable in the NSP
(i.e., stationary) setting. Indeed, once we
decide to admit the time-dependent metrics $\Theta(t)$,
the ket-vector $|\psi(t)\kt \in {\cal H}^{(F)}$
does not suffice to specify, by itself, the state of the system in
${\cal H}^{(F)}$.
For the reasons explained in the related literature \cite{Brody}
one must add also the information about the metric or, equivalently,
about the second ket-vector $|\psi(t)\kkt \in {\cal H}^{(F)}$,
with its time-evolution controlled by the other, conjugate,
independent Schr\"{o}dingerian
Eq.~(\ref{d2.3}).

In a more formal language of review \cite{NIP},
a compatibility
of the probabilistic predictions
(based on formula (\ref{redysMEAS}))
with the
concept of the description of
the state
can most easily be achieved when we represent the states by
elementary projectors
 \be
 \pi_\Theta(t)=|\psi(t)\kt\,
 \frac{1}{\bbr\psi(t)|\psi(t)\kt}\,
 \bbr\psi(t)|
 \label{proopro}
 \ee
i.e., when we decide to
work with the concept of a biorthogonal basis \cite{Brody}
and when we make use of
the knowledge of {\em both\,} the kets $|\psi(t)\kt$
and the (metric-dependent) bras $\bbr\psi(t)|$.

In the complete and physical (hypothetical) Hilbert space ${\cal
H}^{(P)}$ the \textcolor{black}{corresponding twins $|\psi(t)\pkt$ and
$\pbr\psi(t)|$ of the respective kets and bras } form, naturally,
just the ``trivial'' Hermitian-conjugate pairs. Nevertheless, once
the whole theory is formulated in its mathematically friendliest
representation in  ${\cal H}^{(F)}$, we must work with the full
projectors (\ref{proopro}) because only these pure-state projectors
keep trace of the information about the metric as encoded in
$\bbr\psi(t)|$, say, via definition (\ref{dumapo}) \cite{SIGMA}.

\textcolor{black}{This being said we are prepared to return, once
more, to Eq.~(\ref{redysMEAS}). The ultimate necessity of evaluation
of the hypothetical textbook matrix elements $\pbr
\psi^{}(t_f)|\mathfrak{q}_{}(t_f) |\psi^{}(t_f) \pkt$ of
experimentalist's interest is transferred there, via the
metric-containing formula
 $
 \br \psi^{}(t_f)|\Theta(t_f)Q_{}(t_f) |\psi^{}(t_f) \kt$,
from the computationally inaccessible Hilbert space ${\cal H}^{(P)}$
to its technically preferred alternative ${\cal H}^{(F)}$.
It is then necessary to keep in mind that the corresponding
operators representing the observables (being it, in our notation,
$\mathfrak{q}_{}(t)$ in  ${\cal H}^{(P)}$
or $Q_{}(t)$ in
${\cal H}^{(F)}$)
are also
non-stationary.}

\textcolor{black}{As long as the mathematical aspects and consequences
of the latter time-dependence of the operators of observables
will be outlined in
the next subsection, let us now only add that
due to the deeply non-Hermitian nature of these operators
(as well as of their energy-representing special case called Hamiltonian)
a truly important advantage is being reached
when the model in question admits an explicit construction of the integrals
of motion.
In the present non-Hermitian setting, in particular,
the completion of a meaningful quantum
theory is often being perceivably facilitated
whenever one manages to construct the so called
Lewis-Riesenfeld
invariants (for a more detailed exposition and exemplification of such an
efficient trick see, e.g., the recent dedicated paper \cite{invariants}).
}

\subsection{Heisenbergian evolution equations for observables\label{koramo}}

In our preceding comment the point was that for the experimental and
predictive purposes we do not need to
solve
the evolution equation controlling the time-dependence of the metric
at all.
Still,
in principle, this equation might still retain
a role in methodical considerations (see, e.g., \cite{Bila}).
Thus, it may make sense to write it, for the sake of
completeness, down. Still, even in this setting we have
a choice between its Coriolis-force-generated
(or, if you wish, Heisenbergian) version
 \be
 {\rm i}\,\frac{\partial}{\partial t} \,
 \Theta(t)= \Theta(t)\,\Sigma(t) - \Sigma^\dagger(T)\, \Theta(t)\,
 \label{identity}
 \ee
and/or its equivalent Schr\"{o}dingerian alternative
 \be
 {\rm i}\,\frac{\partial}{\partial t} \,
 \Theta(t)= G^\dagger(t)\, \Theta(t)-\Theta(t)\,G(t) \,.
 \label{2.6}
 \ee
One might only add that
a serendipitious merit of the Coriolis-force-controlled approach
may be seen in the fact that the knowledge of $\Sigma(t)$
may be also used in relation
 \be
 {\rm i} \frac{\partial}{\partial t} \,\Omega^{(NIP)}(t)\kt=
\Omega^{(NIP)}(t)\,\Sigma^{(NIP)}(t)\,,
  \label{reniSEFip}
 \ee
i.e., if needed, for a reconstruction of the time-dependent Dyson map.

In the next step let us now return to definition (\ref{redadag})
of a generic observable, and let us differentiate it formally
with respect to time.
What we obtain is a Heisenbergian evolution
equation
 \be
 {\rm i\,}\frac{\partial}{\partial t} \,{Q}_{}(t)=
  Q(t)\,\Sigma(t) -\Sigma_{}(t)\,Q(t)
 +K(t)\,,
 \ \ \ \ \ K(t)=\Omega_{}^{(-1)}(t)\,
 {\rm i\,}\dot{\mathfrak{q}}_{(SP)}(t)\,\Omega_{}(t)\,.
 \label{beda}
 \ee
Although such an equation is an immediate parallel of its
Hermitian interaction-picture (IP) predecessor known in the conventional
quantum theory of textbooks, such a general version of the equation
is
usually considered overcomplicated and
rarely used and solved. In practice,
such an equation is
only
considered usable in
the quantum systems
in which the SP version of the operator remains time-independent, with
$\dot{\mathfrak{q}}_{(SP)}(t)=0$ and, hence, with $K(t)=0$.

Sometimes, an exception is tolerated in non-conservative
systems in which
one may be forced to work, say, with
the
energy-representing Hamiltonian $H(t)$ for which
the
operator
 $$
K^{(NIP)}(t)=\Omega_{}^{(-1)}(t)\, {\rm
i\,}\dot{\mathfrak{h}}_{(SP)}(t)\,\Omega_{}(t)\,
 $$
does not vanish. In such a case the only choice is between the
Heisenbergian version of the Coriolis-controlled equation
 \be
 {\rm i\,}\frac{\partial}{\partial t} \,{H}^{(NIP)}(t)=
  H^{(NIP)}(t)\,\Sigma^{(NIP)}(t)
  -\Sigma^{(NIP)}(t)\,H^{(NIP)}(t)+K^{(NIP)}(t)
 \label{bedakat}
 \ee
and its equivalent, \textcolor{black}{$G(t)-$generated} alternative
 \be
 {\rm i\,}\frac{\partial}{\partial t} \,{H}^{(NIP)}(t)=
     G^{(NIP)}(t)\, H^{(NIP)}(t)
     -H^{(NIP)}(t)\,G^{(NIP)}(t)+K^{(NIP)}(t)\,
 \label{rebedakat}
 \ee
in \textcolor{black}{the derivation of} which one takes the advantage
of the equal-time commutativity of $H(t)$ with itself.


\section{Time-dependent wrong-sign oscillators\label{ifutur}}

\subsection{The Fring's and Tenney's construction\label{futur}}

In the Fring's and Tenney's pioneering application \cite{FT}
of the
NIP approach to
non-stationary wrong-sign oscillators
the authors decided to start from a suitable four-parametric
Ansatz
\begin{equation}
 \Omega^{(FT)}(t)=e^{\alpha (t)x}e^{\beta (t)p^{3}
 +{\rm i}\gamma (t)p^{2}+{\rm i}\delta(t)p}\,.
  \label{veta}
\end{equation}
Such a preselected form of the Dyson map
depended on the four real
and sufficiently smooth functions $\alpha(t) ,\beta(t) ,\gamma(t)$ and $\delta(t)$
of time
(cf. equation Nr. (2.3) in {\it loc. cit.}).
This immediately led to the Coriolis force
 \be
 \Sigma^{(FT)}(t) ={\rm i}x\dot{\alpha}+{\rm i}\dot{\beta}p^{3}
 -(3\dot{\alpha}
 \beta +\dot{\gamma})p^{2}-(2{\rm i}\gamma \dot{\alpha}+\dot{\delta})p
 -{\rm i}\delta \dot{\alpha}
 \label{sigt}
 \ee
where the arguments (time) have been dropped, and where the authors
marked the partial derivatives with respect to time
by an overdot.
In a way inspired by Jones and Mateo \cite{JM}
these authors also picked up the
Schr\"{o}dinngerian generator
 \be
 G^{(FT)}(t)=p^{2}+\frac{m(t)}{4}z^{2}-\frac{\lambda^2(t)}{16}z^{4}\,,
 \ \ \ \ z=z(x)=-2{\rm i}\sqrt{1+{\rm i}x}\,,
 \ \ \ \ x \in \mathbb{R}
 \label{get}
 \ee
defined
in terms of the other two real functions $m(t)$ and $\gamma(t)$ of time.
This enabled them to construct the Hamiltonian-operator (i.e., energy-operator)
 \be
 H^{(FT)}(t)=G^{(FT)}(t)+\Sigma^{(FT)}(t)
 \label{hat}
 \ee
and also the product
$\Theta(t)=\Omega^\dagger(t)\,\Omega(t)$ and the product $\mathfrak{h}(t)=
\Omega(t)\,H(t)\,\Omega(^{-1}t)$.
As long as the sum (\ref{hat}) had to be
$\Theta(t)-$quasi-Hermitian
(i.e., equivalently, as long as the product $\mathfrak{h}(t)$
had to be Hermitian), the authors
just had to
satisfy these conditions via a judicious
adaptation of the parameters.

The result had the form of definition of all of the four
Dyson-map ``output'' functions
$\alpha(t)$, $\beta(t)$ $\gamma(t)$ and $\delta(t)$
in terms of the ``input mass'' $m(t)$, ``input coupling'' $\lambda(t)$
and another real constant $c_1$
(cf. equation Nr. (2.11) in {\it loc. cit.}).
This definition appeared accompanied by another constraint
which has been given its final form of definition of mass and coupling
in terms of a new unconstrained function $\sigma(t)$ of time,
 \be
 \lambda^2(t)=\frac{1}{4\sigma ^{3}(t)},\ \ \ \ \
 \ \ \ \
 m(t)=\frac{4c_{2}+\dot{\sigma}^{2}(t)
 -2\sigma(t) \ddot{\sigma}(t)}{4\sigma ^{2}(t)}\,.
 \label{[2.13]}
 \ee
The symbol $c_2$ denotes here another free real constant
(cf. equation Nr. (2.13) in {\it loc. cit.}).

One can conclude that the simultaneous choice of the
Schr\"{o}dingerian generator $G(t)$
and of the Heisenbergian generator $\Sigma(t)$
led to a feasible construction of the model in which
the authors were able to achieve the ultimate goal of the
guarantee of
isospectrality
between the time-dependent  pre-selected operator sum
$H(t)=G(t)+\Sigma(t)$
and a (still fairly complicated)
self-adjoint operator
$\mathfrak{h}(t)$.

A further,
unitary-transformation-mediated
simplification appeared also feasible.
Although it still led to
a fairly broad multiparameetric family of
non-stationary analogues
of the Jones's and Mateo's tilded stationary Hamiltonian
(\ref{ouha}), a further special choices of parameters
resulted, finally, in a model exhibiting
an almost complete analogy with the Jones's and Mateo's double-well
model (cf. the truly impressive picture Nr. 1 in \cite{FT}).

\subsection{Physical background}

The feasibility of the Fring's and Tenney's closed-form construction
offers a deep insight in the structure of the theory and
reopens the further fundamental questions
concerning the specification
of the physical content of the wrong-sign quantum models
or, in general, in the words of review \cite{Geyer},
of the way we are ``given [a] set of non-Hermitian observables''.

The question has been addressed,
from different points of view, in multiple reviews
(take just \cite{Carlbook,Christodoulides} for illustration)
but the problem may still be considered open.
For the purposes
of its present analysis
it is sufficient to distinguish
between a very broad context of the so called ``non-Hermitian quantum physics''
(in which, in a way sampled by monograph \cite{Nimrod},
the unitarity is not an issue), and the narrower domain of research
(of our present interest)
which could be called the ``quasi-Hermitian quantum physics''.

In the latter setting, naturally, the main emphasis is to be put on the
reality of the spectrum. In this sense, a key conceptual role is played
by the ``implicit''
isospectrality relations given by Eq.~(\ref{dadag}) for
a generic observable, or by
Eq.~(\ref{udagobs}) for the
instantaneous-energy-representing Hamiltonians $H(t)$.
The technical and physical background of the quasi-Hermitian
theory may be then seen to lie in the replacement of the correct
physical Hilbert space
${\cal H}^{(P)}$ (which is, by our basic assumption,
user-unfriendly) by its representation (using the {\it ad hoc\,} metric
operator) in an auxiliary, manifestly unphysical
``mathematical'' Hilbert space
${\cal H}^{(F)}$.

Formally, this means that given an observable $Q(t)$ or $H(t)$
(or, in an extreme setup,
an irreducible set of observables \cite{Geyer}
which are, in general, introduced as non-Hermitian in
${\cal H}^{(F)}$),
the main task is to guarantee the
necessary reality of their spectra.
In this sense, Fring and Tenney
only achieved
the goal by ``brute force''.
{\it A priori}, there was really no reason to believe that
after their ``input'' choice of $\Sigma(t)$ (cf. Eq.~(\ref{sigt}) above)
and of $G(t)$ (cf. Eq.~(\ref{get}) above)
they might eventually succeed in
forcing the spectrum of their
manifestly non-Hermitian Hamiltonian (\ref{hat}) to be real.
That's why we returned to the problem
of the wrong-sign oscillators in our present paper.

An independent word of warning also comes
from paper \cite{NHeisenberg} in which the NSP and NIP formulations
of the quasi-Hermitian quantum mechanics were complemented
by the introduction of non-Hermitian Heisenberg picture (NHP).
In it one simply takes the general NIP formalism and sets,
irrespectively of any dynamics, $G(t) = G^{(NHP)}(t) \equiv 0$.
Naturally, the resulting NHP theoretical framework still admits
an arbitrary specification of the unitary quantum dynamics
(via a suitable
definition of $H(t)$). Nevertheless,
a rather unpleasant surprise was that
the formalism only becomes consistent
in the stationary case, i.e., for the models using
the time-independent metric $\Theta(t)= \Theta(0)=\Theta$.
In this sense, the description of the non-stationary non-Hermitian
quantum systems requires, in general
the use of the full-fledged NIP formalism.
The applicability of
both of its simplified (viz., NSP and NHP) alternatives
remains restricted to the
stationary systems.

\subsection{Alternative, physics-motivated NIP constructions}

In the non-stationary
models which are non-Hermitian in ${\cal H}^{(F)}$
one should resist the temptation of choosing the
``false'' Schr\"{o}dingerian Hamiltonian $G(t)$ or the
Heisenbergian Hamiltonian $\Sigma(t)$
in advance. Indeed,
besides a few mathematical but more or less
purely formal advantages of such an option
(as discussed, more thoroughly, in  \cite{Bila,FringMou,Andreas,ju22} or
in review \cite{NIP}), the
phenomenological role
of these operators
is secondary.
In the manner explained in the older review \cite{Geyer}
the initial physical information about quantum dynamics
is almost exclusively carried by the operators of observables.

In
spite of a certain mathematical user-friendliness
of the choices of $G(t)$ and/or $\Sigma(t)$,
one can only speak about the closed-system
{\em unitary\,} evolution when one manages to guarantee
the reality of the spectrum of the observable of interest
and, in particular, of $H(t)$,
i.e., of {\em the sum\,} of $G(t)$ and $\Sigma(t)$.
In a way illustrated by an
explicitly solvable example
in \cite{3by3},
a consistent NIP description of unitary
systems can be obtained even when the
spectra of $G(t)$ and/or of $\Sigma(t)$
happen to be complex.
Also in the quasi-Hermitian quantum mechanics
of the wrong-sign oscillators
these spectra cannot have any immediate physical
meaning.

An intuitive clarification
of the apparent paradox
is not too
difficult. It is sufficient to imagine that
from the point of view of the predictions of the theory
(mediated by the matrix elements (\ref{redysMEAS}))
the reality of the spectrum of the observable Hamiltonian $H(t)$
is caused by a mutual fine-tuned cancelation of the
complexities generated by the two ``false''
Hamiltonians $G(t)$ and $\Sigma(t)$.

Naturally, whenever one starts from an initial choice of
$G(t)$ and/or $\Sigma(t)$,
a guarantee of the latter cancelation
need not be easy.
This is well documented even by the
Fring's and Tenney's paper \cite{FT} itself.
In it the authors
revealed that in the massless case with $m(t)=0$
one has to fix
 $$
  \sigma^{[0]}(t)=\kappa_0+\kappa_1t+\kappa_2t^2\,,
 \ \ \ \ c_2^{[0]}=\kappa_1\kappa_3-\kappa_2^2/4\,
 $$
due to the second relation in~(\ref{[2.13]}).
Nevertheless,
the Jones' and Mateo's time-independent solutions
of paper \cite{JM} still cannot be reproduced
in any {\it ad hoc\,} limit
because, citing the Fring's and Tenney's own words,
 ``the constraints for $\gamma$ and $\delta$
[in ansatz (\ref{veta}) above]
are different from those reported'' [i.e., reported in equation Nr. (2.11)
of {\it loc. cit.}]
so that
``there is no time-dependent solution corresponding to that choice''
\cite{FT}.

From our present point of view
the resolution of the ``massless-case'' puzzle would lie in the replacement of
the false non-stationary Hamiltonian
$G(t)$ (i.e., of ansatz~(\ref{get}))
by an alternative
definition
 \be
 H^{(NJM)}(t)=-\frac{d^2}{dz^2} - g(t)\,z^4\,
 \label{hahab}
 \ee
of an analogous non-stationary observable JM-like Hamiltonian living
on the same complex contour (\ref{kont}) of the ``coordinate'' $z$
as its stationary predecessor (\ref{haha}).

In such an alternative setting one immediately sees, first of all, that
the innovated, non-stationary wrong-sign NJM oscillator will still possess the
real spectrum. In other words,
the brute-force introduction of the time-dependence
cannot destroy
the observability property of Hamiltonian (\ref{hahab}).
The guarantee of its
spectral equivalence
with its self-adjoint
avatar
 \be
 \widetilde{\mathfrak{h}}^{(NJM)}(t)=
 -\frac{d^2}{dy^2} - 2\,\sqrt{g(t)}\,y
 +4\,g(t)\,y^4\,.
 \label{bouha}
 \ee
as prescribed by Eq.~(\ref{ouha}) is
purely algebraic. The
newly introduced variability with time does not play any role.
One reveals that
in many specific models and applications
the
introduction of a nontrivial and sufficiently general
time-dependence
{\em directly into the observables} (like, e.g., into $H(t)$)
might be not only
warmly phenomenologically welcome (say, in the manybody
quantum physics \cite{Bishop}) but also, in the strictly
mathematical sense, simpler than the
(hardly phenomenologically motivated)
choice of $G(t)$ and/or $\Sigma(t)$.

From such a perspective many transitions to non-stationary dynamical regime
might very easily be realized
by the introduction of a more or less arbitrary
time-dependence
in the original parameters of stationary dynamics which specify, say,
directly the observable Hamiltonian $H(t)$.
In comparison with the
constructions based on the input knowledge of
$\Omega(t)$ and $G(t)$
the alternative input knowledge of the mere $H(t)$
might enable one to
reconstruct, in the manner used by Jones and Mateo \cite{JM},
the admissible forms of $\Omega(t)$ via
the direct factorization of the metric,
i.e.,
more easily, in principle at least.

In the specific and
most elementary Jones' and Mateo's
one-parametric and stationary wrong-sign-oscillator
(\ref{haha}) the elementary replacement of the positive constant $\lambda^2$
by a suitable positive and not too wild time-dependent function $g(t)$
immediately leads to the explicit form of the non-stationary Dyson map
 \be
 \Omega^{(NJM)}(t)= \exp \left [
 p^3/[96\,g(t)]-p
 \right ]
 \label{[19b]}
 \ee
(cf. also the corresponding exact form of the operator of
metric as given in Eq.~({\ref{[19]}}) above).
In the light of this observation
the core of our present message
can be now formulated as a statement that
once one knows the
details of the construction of an arbitrary NSP
model, the knowledge of the stationary version (\ref{dadag}) of the
isospectrality
between the different representations of an
observable in question could often be most
easily extended to its non-stationary NIP version (i.e.,
say, to relation~(\ref{udagobs}) between
Hamiltonians).

The non-stationarity of the operators only enters the game
when one further proceeds to the definitions of
the Coriolis force $\Sigma(t)$
(using the elementary definition (\ref{defsig}), which is
still an entirely straightforward mathematical operation) and
of the related Schr\"{o}dingerian generator $G(t)=H(t)-\Sigma(t)$.
In other words, the price for the easiness of
construction of the non-stationary JM-type Dyson maps
(sampled by Eq.~(\ref{[19b]}))
is to be paid, in applications, by the more complicated forms of the
Heisenberg and Schr\"{o}dinger equations
of the general methodical section \ref{tstatek} above.

\section{Conclusions\label{sekcetreti}}

In the literature, multiple ordinary differential non-Hermitian models
have been studied and described in the stationary NSP framework
(for illustration see, e.g., the L\'{e}vai's chapter in
monograph \cite{Carlbook}).
In parallel, it seems much more difficult to overcome
the technical obstacles encountered in
the more general non-stationary NIP setting.
For this reason, our attention has been attracted by
one of very few
non-stationary and non-Hermitian
wrong-sign models, the exact solvability of which
has been achieved
by Fring and Tenney \cite{FT}.

The appeal of their model as well as of their
method may be found rooted in mathematics as well as in physics.
In both of these contexts
the puzzling ``wrong-sign`` quartic-oscillator
potentials $V(x) = -\lambda^2\,x^4 +{\cal O}(x^3)$
emerged as a byproduct
of analysis of the standard self-adjoint
anharmonic-oscillator Hamiltonian (\ref{ahosc}).
The history of the anomaly has  briefly been recollected by Buslaev
with Grecchi \cite{BG}.
Their careful mathematical study of the
realistic and centrally symmetric $d-$dimensional model (\ref{ahosc})
revealed that
at a fixed
angular momentum quantum number $\ell$,
its bound-state
spectrum of energies
$\{E_{n,\ell}(\lambda)\}$
appears to coincide
with the spectrum of another, complex and non-Hermitian
and, seemingly, far from realistic
Hamiltonian-like
differential operator (\ref{tenles}).

In applied quantum physics, the
references to the $d-$dimensional oscillators of Eq.~(\ref{ahosc})
abound. They
range from the pragmatic constructions of bound states in
atomic and molecular systems \cite{Fluegge} up to the purely
methodical considerations with relevance in field theory
\cite{Witten,Wittenb}.
Their study became also popular
due to its mathematical user-friendliness.
One of the
best known features of the model is that
it offers a sufficiently feasible
tool for the tests and applications of
perturbation theory \cite{Kato,BW,Turbiner}
as well as of the non-Hermitian versions of quantum mechanics \cite{Guardiola}.

In the latter case one really has to be impressed by
the Buslaev's anbd Grecchi's isospectrality results
as mentioned in Introduction. The more so
because
the tilded operator (\ref{tenles}) itself can
be perceived as a
specific quantum Hamiltonian in which just the potential
is complex and asymptotically decreasing.
For such a reason
the discreteness, reality and boundedness
of the spectrum of operator (\ref{tenles}) may
look counterintuitive.

With time,
the
latter
complex
differential Hamiltonian-like operator
appeared to belong to a broader
class with similar properties.
The reality of eigenvalues
has been shown to survive
the omission of the
asymptotically dominant quartic term $-\lambda^2\,x^4$
from interaction
in Eq.~(\ref{tenles})
(see, e.g., the detailed studies of the resulting
imaginary cubic anharmonic oscillators
in \cite{Guardiola,Caliceti,Alvarez}).
The manifestly non-Hermitian models
with real spectra as sampled by Eq.~(\ref{tenles})
were returned to the mainstream of quantum physics.

In our present paper we paid attention to
several formal as well as interpretation
challenges encountered after transition of the
non-stationary version of the theory.
After such a transition
the necessity of making the manifestly non-Hermitian
model compatible with the standard
probabilistic interpretation and postulates
of quantum mechanics appears particularly difficult.
We believe that we managed to clarify a few misunderstandings
which emerged recently in such a context. In this sense, our main message
is that
once one sufficiently clearly distinguishes between the
observable and non-observable operators (which could be all called
Hamiltonians),
the NIP-based treatment of the wrong-sign quartic oscillators
becomes fully acceptable
and void of counterintuitive features and paradoxes.

\end{document}